# Towards the 5th Generation
# of Wireless Communication Systems

## Nicola Marchetti

CTVR / The Telecommunications Research Centre
Trinity College Dublin, Ireland

marchetn@tcd.ie

### Abstract

In this article, one first introduces the general landscape of the next generation of wireless communication systems (5G), including its driver and requirements, and the candidate technologies that might help to achieve its intended goals. The following areas, which the author considers to be of particular relevance for 5G, are then introduced: detection of and access to free spectrum over bands of an heterogeneous nature, extreme densification of networks (massive base station deployments), extreme increase in the number of antennas in transmitter arrays and their interaction with a novel waveform, integration of both wireless and optical sides of telecom networks, and study of wireless networks using the magnifying glass provided by complex systems science. In particular, recent results from the author's research team are shortly discussed for each of these research areas.

## I.    INTRODUCTION

Ongoing and future societal development will lead to changes in the way communication systems are used. On-demand information and entertainment will increasingly be delivered over mobile and wireless communication systems. These developments will lead to a big rise of mobile and wireless traffic volume, predicted to increase a thousand-fold over the next years [1][2]. It is also predicted that today's dominating scenarios of human-centric communication will be complemented by a huge increase in the numbers of communicating machines; there are forecasts of a total of 50 billion connected devices by 2020 [3]. The *coexistence of human-centric and machine-type applications* will lead to a large diversity of communication characteristics.

Different applications will impose very diverse requirements on mobile and wireless communication systems that the fifth generation (5G) will have to support [4]: very stringent latency and reliability requirements, a wide range of data rates, network scalability and flexibility, very low complexity and very long battery lifetime. One of the main challenges is to satisfy these requirements while at the same time addressing the growing cost pressure.

We have fresh quantitative evidence that the wireless data explosion is real and will continue.

A recent Visual Networking Index (VNI) report [5] makes it clear that an incremental approach to 5G will not be enough to meet the demands that networks will face in the next years; indeed likely 5G will have to be a paradigm shift that includes (among other things) very high carrier frequencies with large bandwidths, extreme base station and device densities, and massive numbers of antennas. The motivation behind chasing spectrum in high frequencies is the scarcity of Radio Frequency (RF) spectra allocated for cellular communications. Cellular frequencies use UHF bands for cellular phones, but these frequency spectra have been used heavily, making it difficult for operators to acquire more. Another challenge is that the deployment of advanced wireless technologies comes with high energy consumption; it has been reported by cellular operators that the energy consumption of Base Stations (BS) contributes to over 70% of their electricity bill [6].

A difference compared to previous generations, is that 5G will also be highly integrative, tying any new air interface and spectrum together with Long Term Evolution (LTE) and WiFi to provide universal high-rate coverage and a seamless user experience [7].

Academia is engaging in large collaborative projects such as METIS [4] and 5GNOW [8], while the industry is driving preliminary 5G standardisation activities. To further strengthen these activities, the public-private partnership for 5G infrastructure recently constituted in Europe [7][9]. Fig. 1 shows at a glance the recent evolution of wireless communication systems standards. It is apparent that 5G encompasses elements that are disruptive as compared to the past.

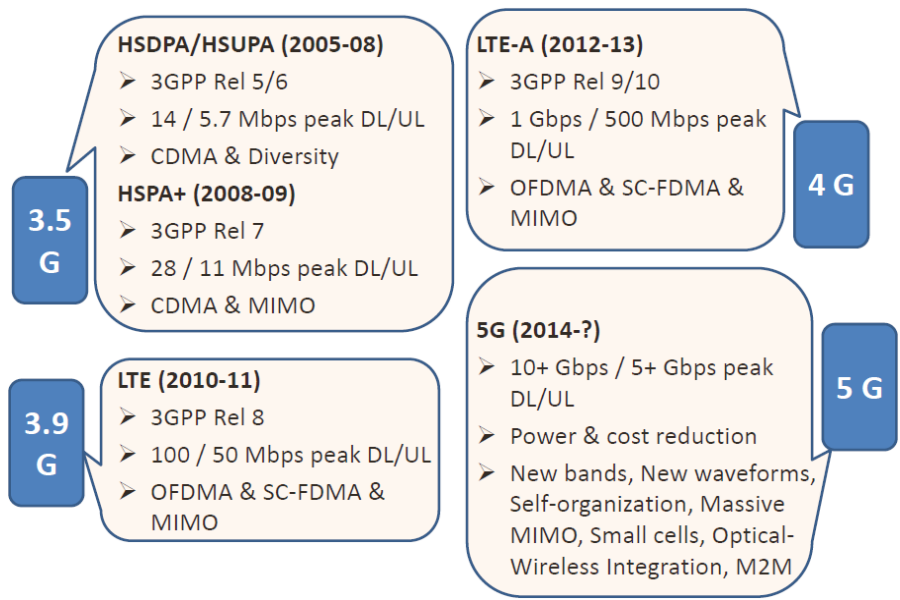

Fig. 1: Some characteristics of recent wireless systems generations.

## II.   DRIVERS AND REQUIREMENTS

While 5G requirements will span every key dimension, not all of them will need to be satisfied simultaneously, as different applications will place different demands on the performance of the systems. For example, very-high-rate applications such as streaming high-definition video may have relaxed latency and reliability requirements compared to e.g. driverless cars or public safety applications, where latency and reliability are paramount but lower data rates can be tolerated [7]. 5G aims to connect the whole planet, achieving seamless and ubiquitous communications between anybody and anything, anywhere, anytime, and anyhow, i.e. by whatever device, service or network [10]. 5G will provide the foundational infrastructure for building smart cities, which will push mobile network performance and capability requirements to their extremes [11]. Among the main *5G drivers*, we have Internet of Things (IoT), Gigabit wireless connectivity and Tactile Internet. For IoT, the main challenge is the scalability problem, with more than 100,000 Machine-to-Machine (M2M) nodes in a cell under the premises of low cost and long lifetime. In terms of Gigabit connectivity, for example users might request quick downloads of 3D streaming content with data rates on the order of 100 Mbps; such levels of connectivity are also expected in large crowd gatherings with possibly interactively connected devices. Tactile Internet comprises a large amount of real-time applications with extremely low latency requirements. Motivated by the tactile sense of the human body, which can distinguish latencies on the order of 1 ms accuracy, 5G aims to be applied to steering and control scenarios, implying a disruptive change with respect to today's communications [12].

As we move to 5G, costs and energy consumption should not increase on a per-link basis. Since the per-link data rates being offered will be increasing by about 100x, this means that the Joules per bit and cost per bit will need to fall by at least 100x; one should therefore try to advocate technological solutions that promise reasonable cost and power scaling. A major cost consideration for 5G, due to the increased BS densities and bandwidth, is how to do the backhauling from the network edges into the core [7].

## III.   CANDIDATE TECHNOLOGY COMPONENTS

A popular view is that the required increase in data rate will be achieved, for the most part, through combined gains in three categories [7]: 1) *extreme network densification* to improve the area spectral efficiency (more nodes per unit area and Hz); 2) *increased bandwidth*, primarily by moving towards mm-wave spectrum and making better use of unlicensed spectrum in the 5 GHz band (more Hz); 3) *increased spectral efficiency*, chiefly through advances in Multiple Input Multiple Output (MIMO) techniques (more bits/s/Hz per node).

As per network densification, its motivation is that a straightforward but extremely effective way to increase the network capacity is to make the cells smaller. Networks are now rapidly evolving to include nested small cells such as picocells (range under 100 meters) and femtocells (WiFi-like range), as well as distributed antenna systems [13][14][15].

In the quest for bandwidth, particular challenges that need to be addressed in 5G systems are fragmented spectrum and spectrum agility. It is unlikely that these challenges can be met using Orthogonal Frequency Division Multiplexing (OFDM), and new waveforms that are more flexible and robust are required. In the 5GNOW project [16] several alternative candidate waveforms are proposed, such as Filter Banks Multi-Carrier (FBMC) [17]. Recent studies suggest that mm-wave frequencies could be used to augment the currently saturated radio spectrum bands for wireless communications [18]. By increasing the RF channel bandwidth for mobile radio channels, the data capacity can be greatly increased, while the latency for digital traffic much decreased, thus supporting much better Internet-based access and applications that require minimal latency. mm-wave frequencies, due to the much smaller wavelength, may exploit new spatial processing techniques such as massive MIMO [19], which can help to achieve larger spectral efficiency.

A native inclusion of M2M communication in 5G involves satisfying three fundamentally different requirements associated with different classes of low data rate services: support of a massive number of low rate devices, sustaining a minimal data rate in virtually all circumstances, and very low latency data transfer [20]. A question for the industry is whether we should have the same network designed for both human and machine communications, a new dedicated network for machines, or a hybrid [21].

It is unlikely that one standard and one model of network deployment will be able to fit all use cases and scenarios in the future. Mobile networks and deployed equipment need to be flexible in order to be optimised for individual scenarios, which may be dynamic in space and time. This requisite for flexibility will have a significant impact on the design of new network architectures. In [22] the authors present one way to provide this flexibility by leveraging cloud technology and exploiting it to operate Radio Access Networks (RAN). Radio access infrastructures based on cloud architecture will provide on-demand resource processing, storage and network capacity wherever needed. Software-defined air interface technologies will be seamlessly integrated into 5G wireless access network architectures, allowing RAN sites to evolve toward a "hyper transceiver" approach to mobile access, helping to realize the joint-layer optimization of radio resources [11].

The ultimate goal of the communication networks is to provide access to information when, where, and in whatever format we need it in. To achieve this goal, wireless and optical technologies play a key role. Wireless and optical access networks can be thought of as complementary. Optical fiber does not go everywhere, but where it does go, it provides a huge amount of available bandwidth. Wireless access networks, on the other hand, potentially go almost everywhere, but provide a highly bandwidth-constrained transmission channel susceptible to various impairments. Future broadband access networks must leverage on both technologies and converge them seamlessly [23].

Fig. 2 summarises some of the *candidate technology components* that the author believes will play a prominent role in the building up and affirmation of 5G. The following sections of the paper will present some highlights about the work done to date in these areas by my group.

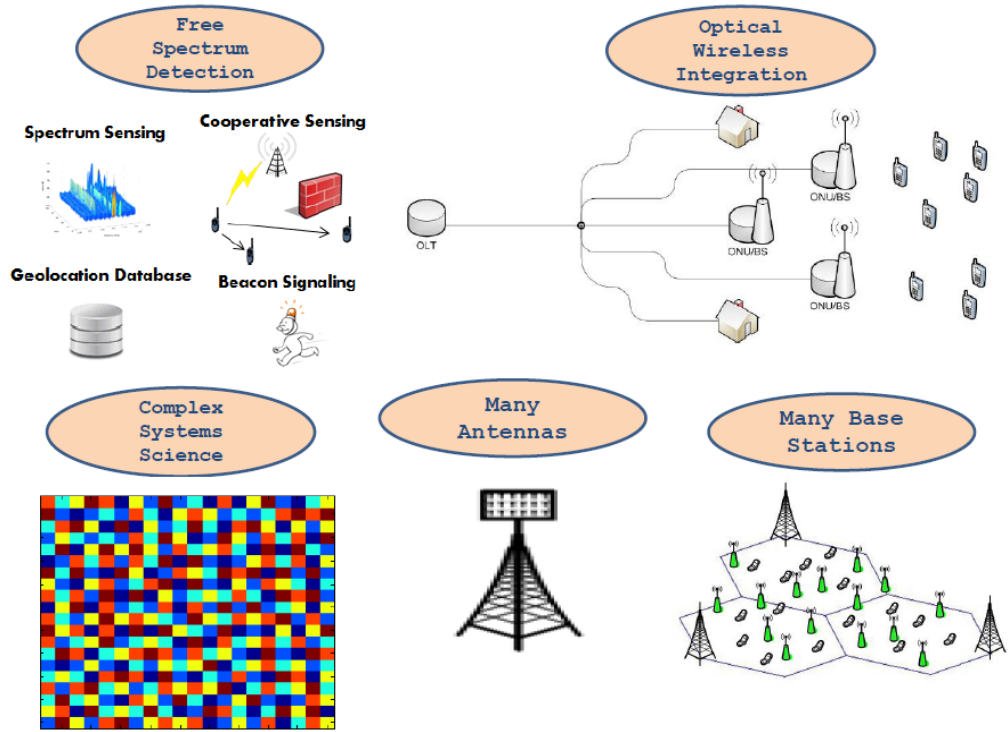

Fig. 2: Some 5G candidate technology components.

## IV.    SPECTRUM

5G systems are expected to provide data rates on the order of Gbps, anytime and anywhere. This can only be realised with much more spectrum than that currently available to International Mobile Telecommunications (IMT) systems through the International Telecommunication Union (ITU) process. All spectrum currently available to cellular mobile systems, including IMT, is concentrated in bands below 6 GHz due to the favourable propagation conditions in such bands. As a result, these bands have become extremely crowded, and prospects for large chunks of new spectrum for IMT systems below 6 GHz are not favourable [24]. As a consequence of this situation, Opportunistic Spectrum Access (OSA) has been considered by regulators for a number of different spectrum bands. In [25] we discuss and qualitatively evaluate techniques used in the discovery of spectrum opportunities, also called white spaces, in the radar, TV, and cellular bands (see Fig. 3). These techniques include *spectrum sensing, cooperative spectrum sensing, geolocation databases*, and the use of *beacons*.

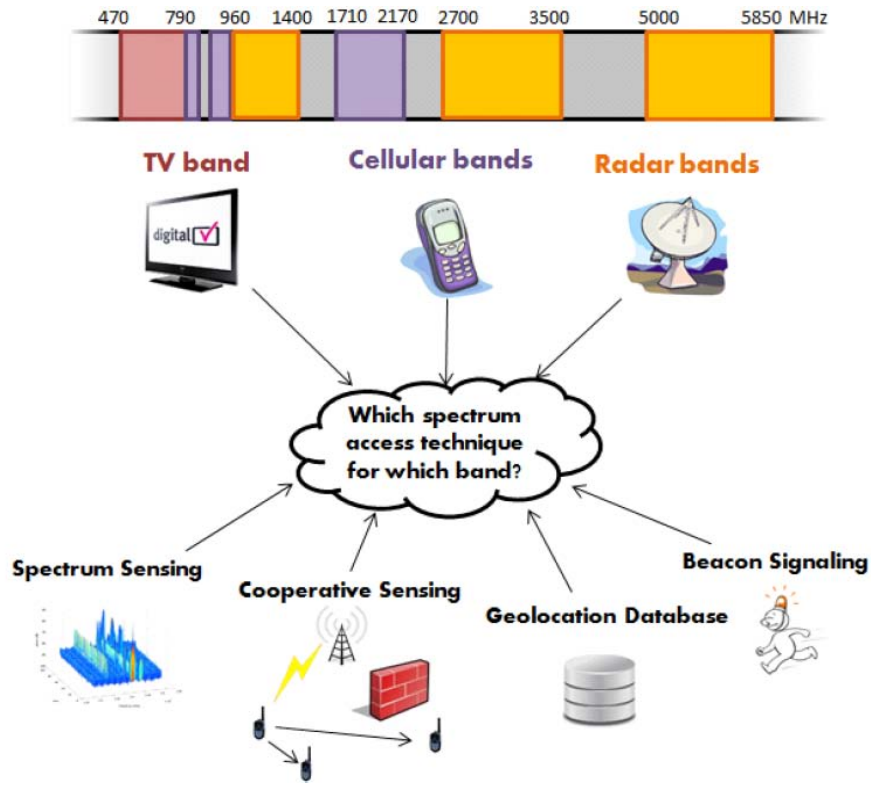

Fig. 3: Which access technique for which band?

Each of the three bands mentioned above, calls for a different set of spectrum access techniques. While TV bands are well suited to the adoption of geolocation databases, a database-assisted spectrum sensing mechanism may represent the most efficient solution to exploit the spectrum holes in radar bands. The unpredictability of cellular systems calls for a more coordinated spectrum access approach, namely beacon signalling, that could be implemented using the already established cellular infrastructure and spare bits of its logical channels [25].

Another potential means to make large amounts of spectrum available to future IMT systems is through Licensed Shared Access (LSA), whereby certain underutilised non-IMT spectrum could be integrated with other IMT spectrum in a licensed pre-determined manner following mutual agreement among the licensees [24]. In [26], we propose a cloud-RAN MD-MIMO (Massive Distributed MIMO) platform as architecture to take advantage of the LSA concept. This particular architecture is worth exploring in the context of LSA since similar principles about the use of resources can be identified. LSA is a spectrum sharing approach to a pool of virtual spectrum resources and cloud-RAN provides the way of managing the *pool of virtual network resources*.

The resources available in a cloud-based RAN (antennas) and LSA (spectrum) are ideally infinite (in practice, much larger than what a single virtual operator requires) but there is a cost associated with their utilisation. Therefore, resource allocation needs to take into account the (unconstrained) pool of resources and their utilisation cost. In [26], we consider the problem of choosing the optimal set of spectrum and antenna resources that maximise the resource efficiency, defined as the number of bits transmitted per cost unit (the resource utilisation cost). In particular, we assume that $K$ users demand a wireless service from a virtual network operator. The operator rents antennas from the cloud-RAN and the LSA spectrum for the time needed to transmit the information. Using the cloud-RAN infrastructure (processing, backhaul, antennas, etc.) and LSA spectrum has a cost, in currency units per second. The aim of the network operator is to choose the optimal number of antennas $M$ and bandwidth $W$ such that the number of transmitted bits per currency unit is maximised (see Fig. 4).

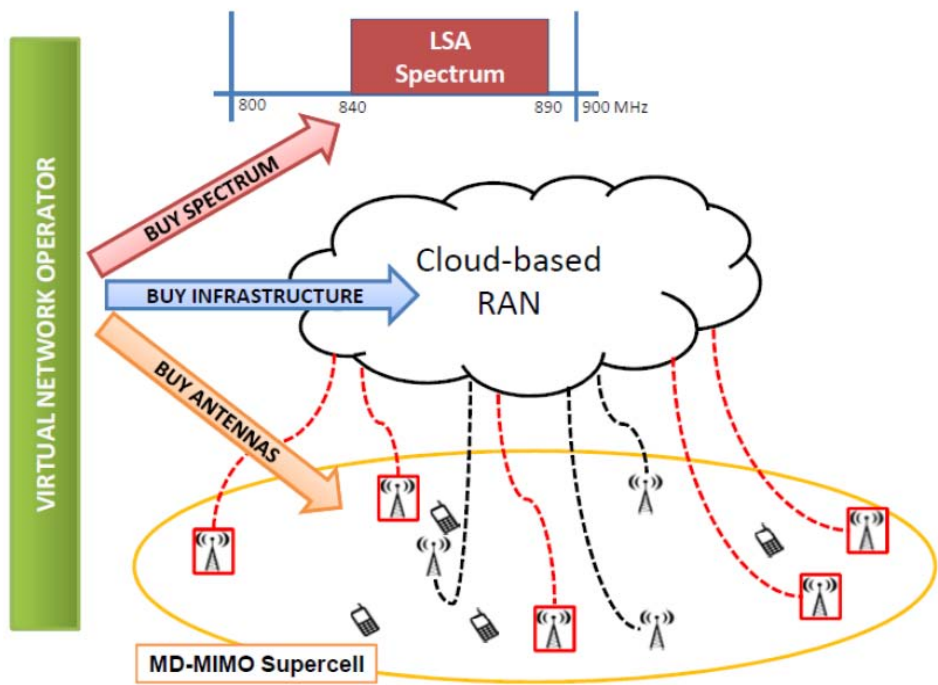

Fig. 4: An example of cloud-based MD-MIMO RAN with $M_{\max} = 8$ and $W_{\max} = 100$ Mhz available antennas and bandwidth, respectively. The virtual network operator uses a subset of the available spectrum ($W = 50$ MHz) and antennas ($M = 5$) to maximise the number of bits transmitted to the $K = 4$ users per cost unit. The red lines indicate the antennas that are being used by the virtual network operator.

The service provider wants to serve $K$ users and chooses $M$ antennas and $W$ MHz from the pool of resources offered by the cloud-based RAN and LSA. Using the infrastructure and the spectrum has a cost, which we identify by the spectrum cost $c_w$ (cost of using 1 MHz of the bandwidth from LSA for 1 second), the antenna cost $c_m$ (cost of using one of the distributed antennas for 1 second), and the operative cost $c_o$ (cost of using the cloud infrastructure, e.g. backhaul, processing, etc. for 1 second).

The cost efficiency is the number of transmitted bits per cost unit (bits/cu). It is computed as the ratio of the total rate to the total costs:

$$\eta(M,W) = \frac{W \sum_{k=1}^{K} \log\left(1 + \frac{r_k}{N_0 W}\right)}{c_m M + c_w W + c_o} \tag{1}$$

where $r_k$ is the power received by the $k$th user and $N_0$ is the noise power spectral density.

Fig. 5 compares the optimal cost efficiency with the efficiency obtained by an arbitrary strategy that either maximises the number of antennas or the bandwidth. We have assumed that $W_{\max} = 50$ MHz and $M_{\max} = 20$. The results show that the optimal solution is able to transmit up to an order of magnitude more information for the same cost. The figure also shows that when the bandwidth cost is very small, maximising the bandwidth is near-optimal. Similarly, if the bandwidth is expensive, the number of active antennas should be maximised.

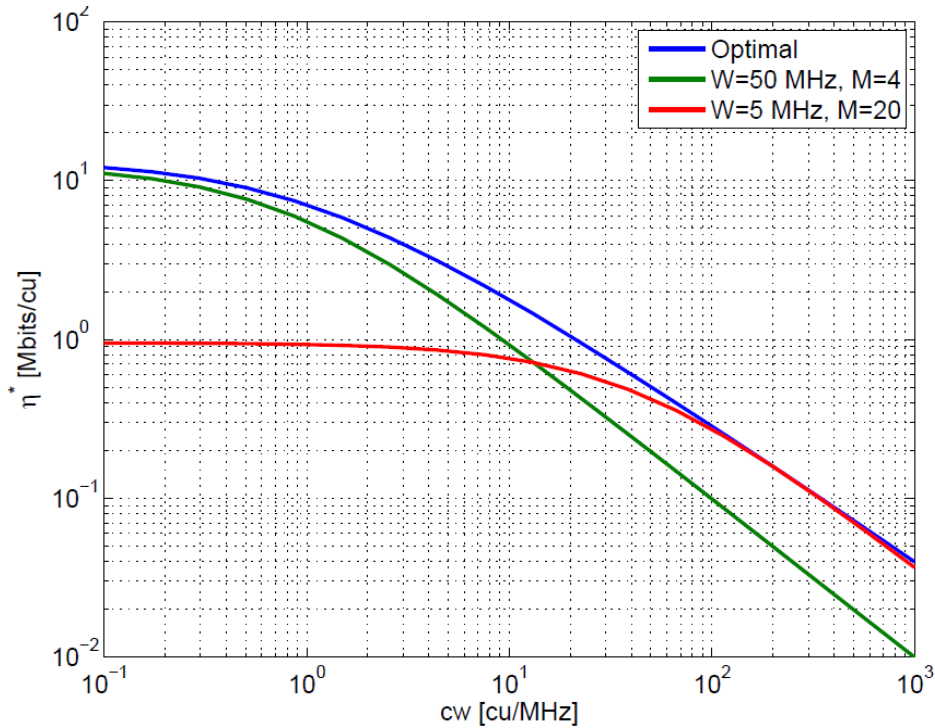

Fig. 5: Cost efficiency when the optimal spectrum and number of antennas are chosen, compared with the cost efficiency when the spectrum or the number of antennas is maximised.

## V. SMALL CELLS

As mentioned in Section III, the number of base stations is expected to grow considerably; there will be more and smaller base stations. This makes it possible to accommodate more users within the same spectrum. However, the base station densification will need to be supported by a widely spread backhaul network. Hence, also the number of backhaul links will increase along with the number of base stations. Backhaul links could be either wired or wireless. Fig. 6 depicts the network densification concept.

The emerging interest in small-cells has pushed researchers to investigate the performance gain of cell splitting. For instance, in dense networks, when the path-loss follows an attenuation proportional to a power of the distance, cell-splitting provides linear Area Spectral Efficiency (ASE) gain with the density of nodes [27]. However the research in small cells has not focused to date on how the total transmit power of the network changes as cell-splitting occurs and what transmit power levels are needed to maintain linear gain. To address this, in [28] we first provide the expression for the minimum transmit power that guarantees linear ASE gain while performing cell-splitting. Then, by applying this expression for the minimum power, we show that the total transmit power of the network (i.e., the sum of the transmit power of all the base stations within a finite portion of the network) needed to achieve linear ASE gain by means of cell-splitting is a decreasing function of the node density, meaning that a significant reduction in the total transmit power can be obtained by shrinking the cell size and increasing the node density.

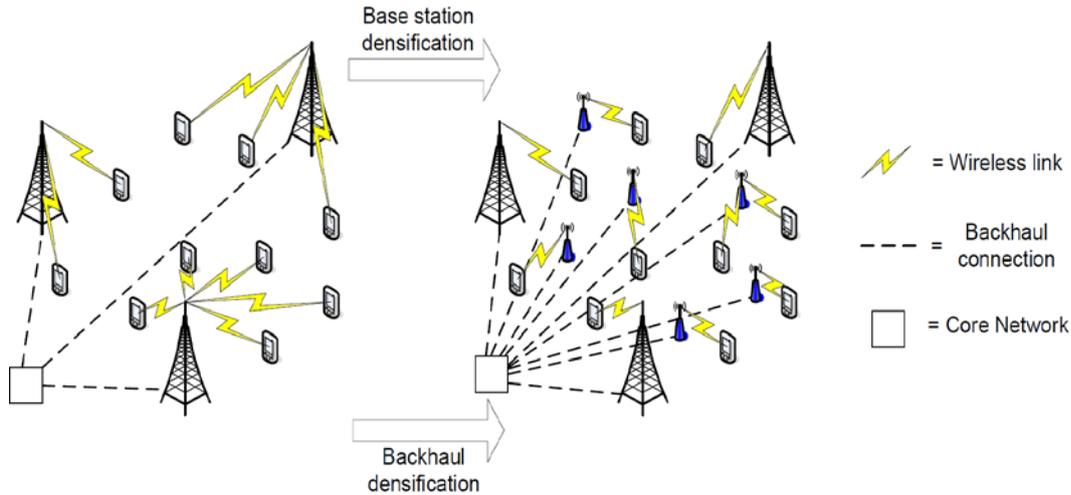

Fig. 6: Network and backhaul densification.

The ASE can be calculated as:

$$\text{ASE} = d \cdot \overline{C_{cell}} \qquad (2)$$

where $d=D^{-2}$ is the cell density, with $D$ being the scaling factor, and $\overline{C_{cell}}$ is the average cell capacity. The total transmit power of the network, obtained by setting the nodes power at the minimum value that still guarantees linear ASE gain when performing network scaling, can be expressed as:

$$P_{TX,tot} = P_0 D^{\beta-2} \overline{\alpha} L^2 \qquad (3)$$

where we consider a finite network portion bounded by a square of area $L^2$ and containing $N$ base stations, $P_0 > 0$ is an arbitrary power, $\beta \in \Re$ is the path loss exponent, $\overline{\alpha} = \dfrac{1}{N}\sum_{k=0}^{N-1}\alpha_k$ and $\alpha_k$ is related to the transmit power of the nth base station by $P_{TX,k} = P_0 D^\beta \alpha_k$.

Fig. 7 shows the *ASE gain and total transmit power reduction*, due to the increase of base station density. As we can see from the plot, the gain is linear with the density of base stations. Nonetheless, the total transmit power used by the network can be decreased while maintaining a linear ASE gain. Hence, the plots show a twofold advantage of increasing the base station density; it allows achieving higher throughput and reducing the overall power radiated by the base station antennas. The overall transmit power reduction achievable by setting the transmit power as indicated in [28], may have positive implications in reducing the aggregate interference experienced by an incumbent willing to share spectrum with a secondary system of small cells. This may be particularly useful in future scenarios involving LSA or Authorised Shared Access (ASA) schemes in which small cell networks exploit new spectrum sharing opportunities.

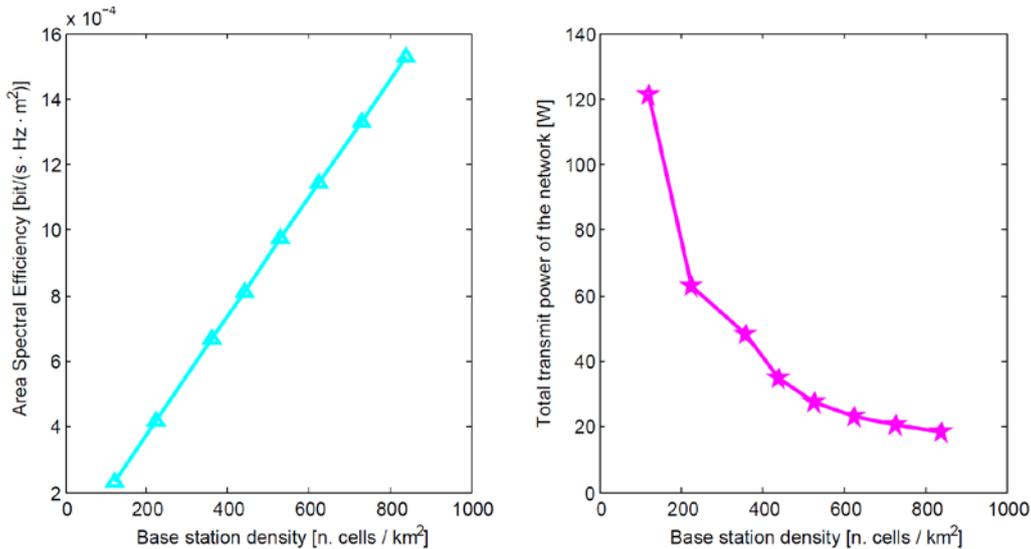

Fig. 7: ASE gain and total transmit power reduction, due to the increase of base station density.

## VI.   COMBINATION OF MASSIVE MIMO WITH FILTER BANK MULTI-CARRIER

In recent years massive MIMO has gained significant momentum as a potential candidate to increase the capacity of multiuser networks, due to the fact that by increasing the number of antennas at BS, the processing gain can be increased to become arbitrarily large. As a consequence, in theory the network capacity can be increased unboundedly [29]. An assumption made by [29] and followed by other researchers is that OFDM may be used to convert the frequency selective channels between each mobile terminal and the multiple antennas at the BS into a set of flat fading channels. Accordingly, the flat gains associated with subcarriers constitute the spreading gain vector that is used for despreading of the respective data stream.

In [30], we introduce the application of FBMC to massive MIMO communications. We find that in the case of massive MIMO systems, linear combining of the signal components from different channels smooths channel distortion; hence one may relax the requirement of having approximately flat gain for the subcarriers, and as a result one may significantly reduce the number of subcarriers in an FBMC system. That way we can reduce both system complexity and the latency/delay caused by the synthesis filter bank (at the transmitter) and the analysis filter bank (at the receiver). Also, one may adopt larger constellation sizes, and hence further improve the system bandwidth efficiency. Moreover, increasing the subcarrier spacing has the obvious benefit of reducing the sensitivity to Carrier Frequency Offset (CFO) and Peak-to-Average Power Ratio (PAPR). An additional benefit of FBMC is that carrier/spectral aggregation (i.e., using non-contiguous bands of spectrum for transmission) becomes a simpler task, since each subcarrier band is confined to an assigned range and has a negligible interference to other bands; this is not the case with OFDM [31].

Fig. 8 presents some results that show the effect on the Signal-to-Interference Ratio (SIR) of increasing the number of antennas at the receiver, for different number of subcarriers in a single-user case, exploring the system capability of achieving a flat channel response over each subcarrier band. For the channel model used here, the total bandwidth, equivalent to the normalised frequency one in the figure, is equal to 2.8 MHz. This, in turn, means the subcarrier spacing in each case is equal to $2800/L$ kHz, where L is the number of subcarrier bands. As an example, when $L = 32$, subcarrier spacing is equal to 87.5 kHz. Compared to the subcarrier spacing in OFDM-based standards (e.g., IEEE 802.16 and LTE), this is relatively broad; $87.5/15 \approx 6$ times larger. Reducing the number of subcarriers (equivalently, increasing symbol rate in each subcarrier band), as noted above, *reduces transmission latency, increases bandwidth efficiency, and reduces sensitivity to CFO and PAPR.*

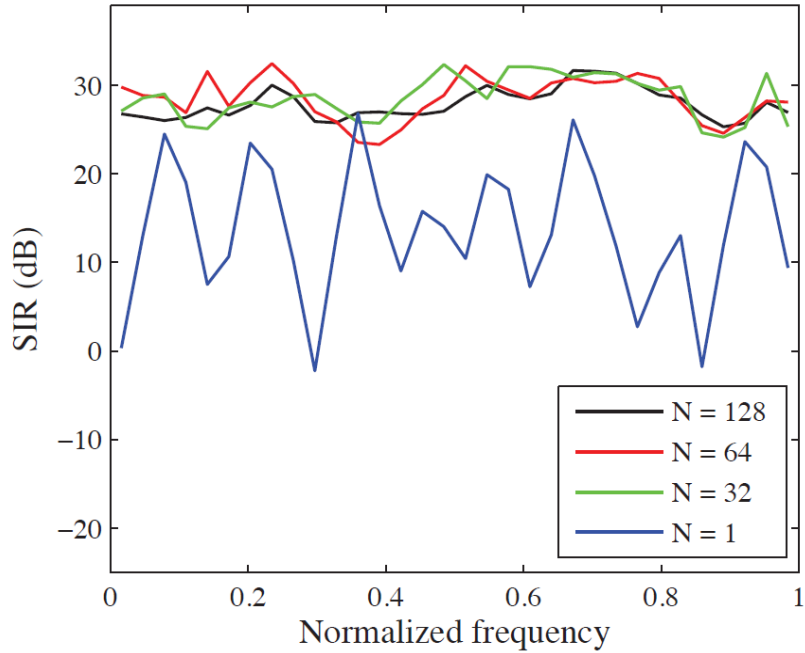

Fig. 8: Comparison of the SIR of matched filter linear combining technique, for different number of receive antennas $N$ and for the cases of 32 subcarriers.

## VII. OPTICAL-WIRELESS INTEGRATION

The main reason why it has taken many years for Fibre To The Home (FTTH) access systems to be deployed is one of financial viability. Ultimately, the main technology that enabled the wide spread of fibre access was the Passive Optical Network (PON), since it allowed sharing the costs of installing fibre and deploying electronic terminations among multiple users [32]. As mobile network operators start looking into the deployment of large numbers of small cells, to offer a higher capacity per user, the idea of a shared, low-cost, fibre backhaul network based on existing PON systems becomes especially attractive. PON systems seem thus an ideal technology for mobile backhaul of small (and large) cells, due to the potentially ubiquitous presence of access points that they can offer. Indeed, the latest ITU-T PON standard, 10-gigabit-capable Passive Optical Network (XG-PON), already considers *Fibre To The Cell* (FTTCell) scenarios [33]. In Fig. 9, we can see an illustration of an FTTCell scenario where an XG-PON network is used to backhaul an LTE system, by connecting the base stations to the core components of LTE. In this architecture, the core components are connected to the root of the PON, also called Optical Line Terminal (OLT) while the base stations are connected to the Optical Network Units (ONU), at the leaves of the tree shaped optical distribution network.

Typically, PONs and their Dynamic Bandwidth Assignment (DBA) algorithms, are designed for ONUs that are independent from each other (often representing individually billed entities). However, scenarios can be envisaged, FTTCell being an example, where wireless operators may require more than one ONU per PON to provide service in different locations. These entities may desire to have a single service level agreement for their group of ONUs and share the contracted capacity among the entire group of ONUs. In [34] we discuss hierarchical DBA algorithms that allow scheduling bandwidth to a group of base stations rather than individually. By assuring bandwidth to a group of base stations, a mobile operator can ensure that whenever one base station is not using its assured bandwidth, it can be assigned first to a base station of the same operator, rather than anyone else in the PON.

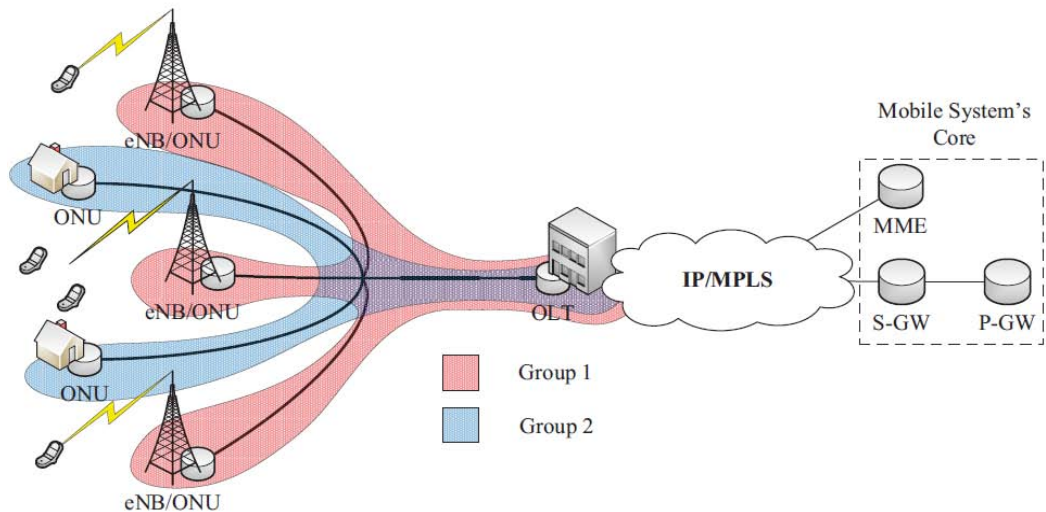

Fig. 9: Fibre To The Cell architecture.

By doing this, mobile operators can make use of the properties of statistical multiplexing, taking advantage of heterogeneity of the traffic from the base stations. With careful dimensioning, it is possible for base stations to transmit at their peak rate, without assuring the peak rate to each base station. While we could argue that the same effect could be achieved with best-effort bandwidth, without assuring bandwidth to the group, the performance of the backhaul would in that case depend on other users on the PON, possibly competing mobile operators.

In [34] we propose group-GIANT (gGIANT), an algorithm developed to enable the assignment of group assured bandwidth, and evaluate its performance through simulations in homogeneous and heterogeneous traffic scenarios. In particular, the heterogeneous simulation consists of changing the load of only one ONU, and keeping constant the load of all the other ONUs. The results for the average delay can be seen in Figure 10, where N indicates the size of the group, i.e. the number of ONUs belonging to it.

We can see that by adding ONUs to the group, extra capacity is available to the ONUs that need it. Comparing the more homogeneous and the more heterogeneous scenarios, we indeed saw in [34] that the results support the idea that gGIANT algorithm provides a larger performance increase when the traffic load is unbalanced. This supports the idea that the more heterogeneous and bursty the traffic is, the bigger the gains from group assured bandwidth that can be obtained.

## VIII.   A COMPLEX SYSTEMS SCIENCE VIEW OF 5G

Modern ICT systems are becoming increasingly larger by englobing more and more components, while at the same time there is an evergrowing flow of information in the systems. As the technological and social trend of communications is shifting from systems based on closed hierarchical or semi-hierarchical structures to open and distributed, networked organisations, new paradigms are needed to model, design, monitor, and control this new kind of systems. Communication engineers are defied by the task of designing networks capable of self-organisation, self-adaptation and self-optimisation of their interactions, and that can satisfy user demand without disruptions. In this regard, help comes from recent studies of complex systems in nature, society and engineering, which suggest ideas on how to design and control modern communication systems [35].

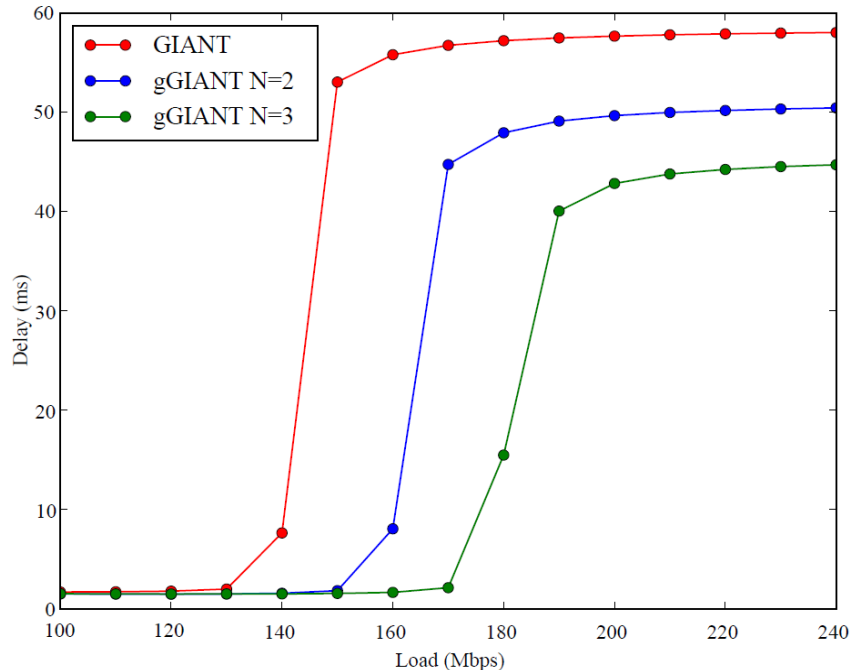

Fig. 10: Average delay of upstream transmission when increasing the load of only one ONU.

In [36] we move a step towards a comprehensive and rigorous study of communication systems, by using understanding and tools from complex systems science. As a specific application of complex system science to wireless systems, we tackle the problem of self-organising frequency allocation, using local information and adaptations to achieve global network-wide behaviour. We support the claim that the system we are considering is a complex one, both in terms of entropy and complexity metrics, and show that simple agents such as cellular automata cells, can actually achieve a nontrivial interference-free frequency allocation.

It is known from complex systems science literature that complexity and entropy are two distinct quantities. For studies on the relation between complexity and information/entropy, readers can refer to [37] [38]. In [36] we use excess entropy to measure complexity. Excess entropy can be expressed in different ways; the form we utilise is the convergence excess entropy EC, which is obtained by considering how the entropy density estimates converge to their asymptotic value h. In two dimensions the entropy density h can be expressed as [39]:

$$h = \lim_{M \to \infty} h(M) \tag{4}$$

where $h(M)$ is the entropy of a target cell $X$ conditioned on the cells in a certain neighbourhood of $X$. Then the excess entropy $E_C$ is defined as:

$$E_C = \sum_{M=1}^{\infty} \big( h(M) - h \big) \tag{5}$$

We then study the global network-wide behaviour with respect to the complexity of the channel allocation matrix. Fig. 11(a) shows the channel allocation matrix resulting from the regularly spaced assignment of $N = 6$ channels. This is a typical example of the frequency allocation resulting from a centralised frequency planning. Fig. 11(b) shows an example of the channel allocation matrix resulting from the self-organising algorithm described in [36]. Finally, a random allocation of $N = 6$ channels is shown in Fig. 11(c).

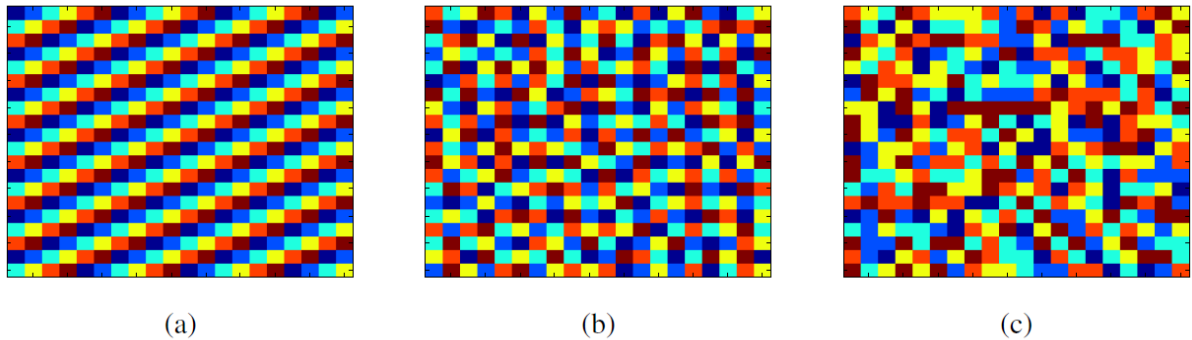

(a)                                    (b)                                    (c)

Fig. 11: (a) Regular channel assignment. (b) Channel assignment resulting from the self-organising algorithm described in [36]. (c) Random channel assignment.

We estimate $E_C$ and h for the three types of channel assignments in Figure 11 by using $10^4 \times 10^4$ matrices. For the channel assignment in Fig. 11(a), the entropy estimates are $h(M) = 0, \forall M = \{1,2,\ldots,6\}$. Hence, $E_C = 0$ and h = 0. This is consistent with the crystal-like completely ordered structure of the channel allocation matrix. For the random channel assignment matrix, the entropy estimates are $h(M) = 2.58, \forall M = \{1,2,\ldots,6\}$. Hence, $E_C = 0$ and $h = 2.58$. As the channel assignment matrix is completely disordered, the entropy is the maximum possible for N = 6 channels, i.e. $\log_2(6)$. Finally, in the case of the channel assignment matrix in Fig. 11(b), the entropy estimates are $h(M) = 1.29, \forall M = \{4,5,6\}$. Hence, $h = 1.29$ and the resulting $E_C$ is 2.04. Therefore the channel assignment emerging from self-organisation exhibits a high amount of *structure* which neither the centralised nor the random channel allocation can reach.

For networks that allocate frequencies in a self-organised manner, we then showed that it makes sense to talk in terms of complex systems. Instead, in the case networks manage frequencies in a centralised fashion, as the resulting allocation would be very much like a regular crystal-like configuration, there is no point in studying them using complex systems science. We consider the work of [36] as a step towards a comprehensive and rigorous study of complex communication systems, adopting in communication networks design and analysis, philosophy and results from the emerging multi-disciplinary field of complex systems science.

## IX.    CONCLUSIONS AND OUTLOOK

It is a time of unprecedented change, where traffic on telecommunication networks is growing exponentially, and many new services and applications are continuously emerging. In this scenario we cannot predict exactly what lies ahead, and therefore the best we can do is to extrapolate some trends and make educated guesses. A possible way out is to design networks with change in mind, so that they will be more robust to disruption caused by growing demands and changing user patterns and yet-unimagined applications, and that the risks associated with investment in these kinds of networks will be lower, as they will be more durable and scalable. Networks that are designed with change in mind will also make effective use of resources (e.g., spectrum, bandwidth, power, processing capabilities, backhaul, etc.) and ensure a sustainable future.

The next generation of wireless communication systems (5G) is part of the above picture. In this article, one introduced the general landscape of 5G systems, including their likely requirements and the candidate technologies promising to achieve such goals. A few relevant 5G areas have been discussed and recent research results in these domains from the author's research team have been shortly presented. In the last part of the article, one focused on the complex systems science's view of future communication networks. One can point out that one of the most widely accepted definitions of complex system, is that of "a system in which large networks of components with no central control and simple rules of operation give rise to complex collective behaviour, sophisticated information processing, and adaptation via learning or evolution" [40]. This view resonates with the author's research team understanding of future wireless networks; indeed networks are becoming increasingly distributed, formed by an ever growing amount of nodes that must take local decisions (due to limits in signalling exchange capacity) reacting to the surrounding environment, and yet have to achieve a global level of good user experience and network performance in general. *There is therefore ground to believe that telecommunication systems are evolving from being simple monolithic structures to complex ones, and that complex systems science might prove beneficial in their analysis and design.*

## ACKNOWLEDGMENT

The author acknowledges support from the Irish CTVR CSET grant 10/CE/I1853.